\documentclass[iopscience,apj]{emulateapj}

\usepackage{placeins}
\usepackage{natbib}
\usepackage{amsmath}
\usepackage{color}

\begin{document}

\newcommand\msun{M_{\odot}}
\newcommand\lsun{L_{\odot}}
\newcommand\msunyr{M_{\odot}\,{\rm yr}^{-1}}
\newcommand\be{\begin{equation}}
\newcommand\en{\end{equation}}
\newcommand\cm{\rm cm}
\newcommand\kms{\rm{\, km \, s^{-1}}}
\newcommand\K{\rm K}
\newcommand\etal{{et al}.\ }
\newcommand\sd{\partial}
\newcommand\mdot{\dot{M}}
\newcommand\rsun{R_{\odot}}
\newcommand\yr{\rm yr}

\shorttitle{PLANETARY SIGNATURES IN SAO 206462} 
\shortauthors{Bae et al.}

\title{PLANETARY SIGNATURES IN THE SAO 206462 (HD 135344B) DISK: A SPIRAL ARM PASSING THROUGH VORTEX?}

\author{Jaehan Bae\altaffilmark{1},
Zhaohuan Zhu\altaffilmark{2,3},
Lee Hartmann\altaffilmark{1}
}

\altaffiltext{1}{Department of Astronomy, University of Michigan, 1085 S. University Avenue,
Ann Arbor, MI 48109, USA} 
\altaffiltext{2}{Department of Astrophysical Sciences, Princeton University,
4 Ivy Lane, Peyton Hall, Princeton, NJ 08544, USA}
\altaffiltext{3}{Hubble Fellow.}

\email{jaehbae@umich.edu, zhuzh@astro.princeton.edu, lhartm@umich.edu}

\begin{abstract}

The disk surrounding SAO 206462, an 8~Myr-old Herbig Ae star, has recently been reported to exhibit spiral arms, an asymmetric dust continuum, and a dust-depleted inner cavity.
By carrying out two-dimensional, two-fluid hydrodynamic calculations, we find that a planetary-mass companion located at the outer disk could be responsible for these observed structures.
In this model, the planet excites primary and secondary arms interior to its orbit.
It also carves a gap and generates a local pressure bump at the inner gap edge where a vortex forms through Rossby wave instability.
The vortex traps radially drifting dust particles, forming a dust-depleted cavity in the inner disk.
We propose that the vortex is responsible for the brightest southwestern peak seen in infrared scattered light and sub-millimeter dust continuum emission.
In particular, it is possible that the scattered light is boosted as one of the spiral arms passes through the high density vortex region, although the vortex alone may be able to explain the peak.
We suggest that a planetary companion with a mass of 10--15~$M_J$ is orbiting SAO~206462 at 100--120~au.
Monitoring of the brightest peak over the next few years will help reveal its origin because the spiral arms and vortex will show distinguishable displacement.

\end{abstract}


\section{INTRODUCTION}

The protoplanetary disk around SAO~206462 is one of the few systems so far to exhibit asymmetric features in both near-infrared scattered light and sub-millimeter thermal emission.
Direct polarimetric imaging revealed two spiral arms in the H and K$_s$ bands \citep{muto12,garufi13}, and the dust continuum emission at sub-millimeter wavelengths observed with ALMA showed a vortex-like feature as well as a dust-depleted inner cavity \citep{perez14,vandermarel15a,vandermarel15b}.
One possibly interesting feature is that the scattered light observations indicate an abrupt change in brightness and pitch angle in one of the arms, while most models show smooth spiral structure \citep[e.g.][]{fung15}. 

Recent studies have shown that planet-driven spiral arms may explain the observed scattered light in near-infrared \citep[e.g.][]{dongetal15,zhu15}.
Upon their formation, planets excite density waves in disks with different azimuthal modes.
While the different modes tend to interfere with each other and merge to a single spiral arm \citep{ogilvie99}, it is found that a secondary can also exist and the separation between these two arms increases with the planet mass \citep[e.g.][]{fung15,zhu15}.
Planets are also capable of generating vortices at the inner and outer gap edge, where pressure has local maxima \citep{koller03,li05,devalborro07,lin10,lyra13,fu14a,fu14b,zhustone14,zhu14}.
The vortices efficiently trap dust particles and impede dust migrating inward, possibly forming a dust-depleted inner cavity.

The scattering of optical and near-infrared light and the thermal sub-millimeter emission reflect not only the gas structure but the spatial distribution of the dust particles with different sizes.  
The small dust responsible for the scattered light should be well-coupled to the gas, but larger grains responsible for millimeter-wave emission can be concentrated by pressure gradients.

In this work, we test whether a planetary-mass companion is able to generate various structures observed in the SAO~206462 disk.
In order to test our models against near infrared scattered light and sub-millimeter thermal emission, we perform two-fluid calculations where we can simulate dust as well as gas structures.
Based on the calculations, we show that the observed structures are well reproduced with a 10--15~$M_J$ planet orbiting SAO~206462 at 100--120~au.
The abrupt change in brightness in one of the arms could be produced as the arm passes through the high density vortex region, though vortex alone may be able to explain the bright scattered light peak. 
We discuss possible future observations that will help reveal the origin of the structures seen in SAO~206462 disk and better constrain planetary mass and position.

\section{NUMERICAL METHODS}
\label{sec:methods}

\subsection{Disk Model}
\label{sec:model}

\begin{deluxetable*}{lcc}
\tablecolumns{3}
\tabletypesize{\scriptsize}
\tablecaption{Model Parameters\label{tab:parameters}}
\tablewidth{0pt}
\tablehead{
Parameter & Value
}
\startdata
Stellar mass ($M_*$) & $1.7~M_\odot$\tablenotemark{a} \\
Disk mass ($M_{\rm disk}$) & $0.026~M_\odot$\tablenotemark{b} \\
Disk aspect ratio at planet's orbit ($(H/R)_p$) & 0.1 \\
Planet mass ($M_p$) & $5, 7.5, 10, 12.5, 15, 17.5, 20~M_J$ \\
Semimajor axis of the planet ($R_p$) & $80, 90, 100, 110, 120, 130, 140$~au \\
Minimum grain size ($a_{\rm min}$) & $0.1~\mu$m \\
Maximum grain size ($a_{\rm max}$) & 5.3~mm 
\enddata
\tablenotetext{a}{\citet{muller11}}
\tablenotetext{b}{\citet{andrews11}}
\end{deluxetable*}

We begin with an initial power-law surface density of the disk gas
\begin{equation}
\Sigma_g (R) = \Sigma_p\left({R \over R_p}\right)^{-1},
\end{equation}
where $\Sigma_g$ is the gas surface density and $\Sigma_p$ is the gas surface density at the semi-major axis of the planetary orbit $R_p$.
The power-law density slope of $-1$ is consistent with the best solution of \citet[][see their model 5]{carmona14} found in between 30 and 200~au.
We choose $\Sigma_p$ such that the initial total gas mass is $0.026~\msun$ \citep{andrews11}.

In these two-dimensional $(R, \phi)$ simulations, we use a fixed radial temperature distribution with the isothermal equation of state, which implies that the ratio of disk scale height to radius is
\begin{equation}
\label{eqn:hr}
{H \over R} = \left( {H \over R}\right)_p \left({R \over R_p}\right)^{0.25},
\end{equation}
where $(H/R)_p$ is the aspect ratio at planet's semi-major axis.
We use $(H/R)_p = 0.1$ which results in the corresponding disk temperature profile 
\begin{equation}
\label{eqn:disk_temperature}
T = 44~{\rm K}\left( {R \over R_p}\right)^{-0.5} \left({R_p \over {100~{\rm au}}}\right)^{-1}
\end{equation}
with $M_*=1.7M_\odot$ and mean molecular weight of 2.4.
The temperature profile is roughly consistent with detailed flared disk models, including good agreement with estimates of the disk scale heights for SAO~206462 from \citet{andrews11} and \citet{carmona14}.
Also, the temperature dependence on radius is consistent with the assumption of a constant $\alpha$ and $\Sigma \propto R^{-1}$ (for a steady disk).

For the gas component a uniform viscosity parameter $\alpha=10^{-4}$ is implemented.
Vortex formation through the Rossby wave instability (RWI) could be dependent on the choice of $\alpha$ value as seen in numerical simulations \citep[e.g.][]{bae15}.
Empirically, it has been shown that the width of the gas pressure maximum has to be $\lesssim 2H$ in order for the RWI to develop \citep{lyra09,regaly12}.
Although determining the critical upper $\alpha$ value for which our results remain valid could be done only numerically given the complexity of the disk structure (planet, spiral arms, vortex, their interaction, etc.), adopting this criterion, one could make a simple argument that the viscous timescale at the inner gap edge ($t_{\nu}$; the timescale that the disk viscously spreads out the gas pressure bump by $\sim H$) has to be longer than the vortex formation timescale.
In an alpha disk, the corresponding viscous timescale can be written as $t_{\nu} = H^2/\nu = P/(2 \pi \alpha)$, where $\nu=\alpha H^2 \Omega$ is the viscosity and $P=2\pi / \Omega$ is the local orbital period.
We find that vortices form at the inner gap edge within 10--20 local orbital time; thus, unless the viscosity parameter is as large as $\sim 0.01$ our results should remain valid.

We adopt inner and outer boundaries at 20 and 300~au. 
We use 288 logarithmically spaced radial grid-cells and 688 linearly spaced azimuthal grid-cells, with which choice $\Delta R /R$ is constant to 0.009 and grid-cells have comparable radial and azimuthal sizes at all radii.

\subsection{Dust Component}
\label{sec:dust}

The dust response to planet-generated gas structures is calculated with the two-fluid FARGO code introduced in \citet{zhu12}, in which the dust component is added to the standard FARGO code \citep{masset00}. 
To briefly summarize, we solve an additional mass and momentum equation set for the dust component.
We treat the dust component as an inviscid, pressureless fluid so dust simply feels the drag force in addition to the central stellar potential.
The drag terms are added to the momentum equation as an additional source step
\be\label{eqn:dragr}
{\partial v_{R,d} \over \partial t} = - {v_{R,d} - v_{R,g} \over t_s}
\en
and 
\be\label{eqn:dragp}
{\partial v_{\phi,d} \over \partial t} = - {v_{\phi,d} - v_{\phi,g} \over t_s},
\en
where $v_R$ and $v_\phi$ are radial and azimuthal velocities, and the subscripts $g$ and $d$ denote gas and dust components.
The dust stopping time is $t_s = {\pi \rho_p a / 2 \Sigma_g \Omega}$ with $\rho_p$ and $a$ being dust particle density and size.

Since dust particles can diffuse in the gaseous disk due to turbulence, dust diffusion is implemented in the operator split fashion in the source step of the dust component \citep{clarke88}:
\be
{\partial \Sigma_{d} \over \partial t} = \nabla \cdot \left( D \Sigma_g \nabla \left({\Sigma_d \over \Sigma_g}\right) \right).
\en
In the above equation, $D = \nu / {\rm Sc}$ is the turbulent diffusivity where $\nu$ is the gas viscosity and ${\rm Sc} = 1 + (\Omega t_s)^2$ is the Schmidt number \citep{youdin07}.

\begin{figure*}
\centering
\epsscale{1.15}
\plotone{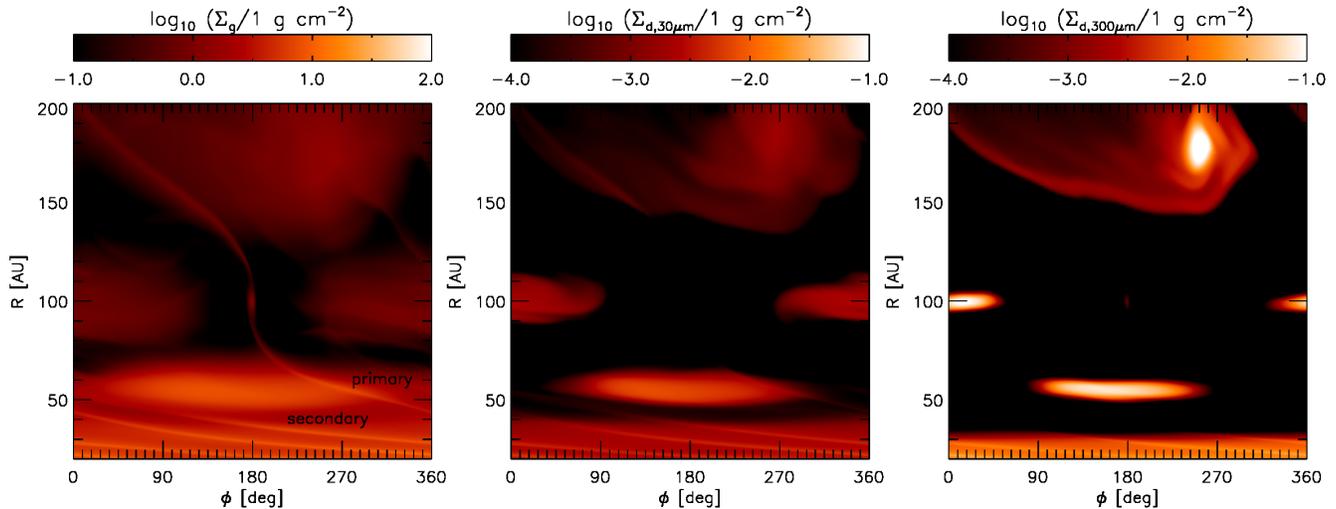}
\caption{Distributions of (left) gas, (middle) $30~\mu m$ dust, and (right) $300~\mu m$ dust at $t=50~T_p$ in $\phi-R$ coordinates. $M_p = 10~M_J$ and $R_p = 100$~AU were used. The planet excites primary and secondary arms at the inner disk. It also generates horseshoe region and opens a gap. Two vortices are formed at the inner and the outer gap edge. The image is shifted in azimuth from its original in a way that the planet locates at $\phi=180^\circ$.}
\label{fig:simulation}
\end{figure*}

We perform calculations with five different dust sizes: 30, 100, 300, 1000, 3000~$\mu$m, which represent particles in size bins of [17~$\mu$m, 53~$\mu$m], [53~$\mu$m, 170~$\mu$m], [170~$\mu$m, 530~$\mu$m], [530~$\mu$m, 1700~$\mu$m], [1700~$\mu$m, 5300~$\mu$m], respectively.
We assume that the dust particles have a power-law size distribution $dn(a) \propto a^{-3.5} da$  in between $a_{\rm min} = 0.1~\mu$m and $a_{\rm max} = 5.3$~mm.
We note that we do not perform calculations for the particles with $a < 17~\mu$m, and simply assume that those particles perfectly follow the gas distribution when producing synthetic ALMA observations in Section \ref{sec:discussion}.
The dust mass in each size bin is determined based on the particle size distribution of $dn(a) \propto a^{-3.5} da$, which provides the surface density as a function of $a$: $d\Sigma_d (a) \propto a^{-0.5} da$.
The mass fractions in each size bin of [0.1~$\mu$m, 17~$\mu$m], [17~$\mu$m, 53~$\mu$m], [53~$\mu$m, 170~$\mu$m], [170~$\mu$m, 530~$\mu$m], [530~$\mu$m, 1700~$\mu$m], [1700~$\mu$m, 5300~$\mu$m], are 5.3~\%, 4.4~\%, 7.9~\%, 13.8~\%, 25.1~\%, 43.6~\%, respectively.
The total initial dust surface density is assumed to be $1~\%$ of the gas density.
We assume that the dust density is $3~{\rm g~cm^{-3}}$.

We vary the planet's mass and semi-major axis to model the SAO~206462 disk.
The model parameters are summarized in Table \ref{tab:parameters}.

\section{RESULTS}
\label{sec:results}

All the calculations were run for $100~T_p$ where $T_p$ refers to the orbital time at the planet's location.
We find that the spiral structure is well developed within $2~T_p$ and remains steady afterward.
The planet carves a gap and generates a horseshoe orbit region around its orbit.
It also generates vortices at the inner and outer gap edges.
The vortex at the inner gap edge forms within $\sim5~T_p$ and survives until the end of the calculations.
The duration of the calculations is thus long enough for the structures to fully develop in the disk.

Figure \ref{fig:simulation} shows gas, 30~$\mu$m dust, and $300~\mu$m dust distributions at $50~T_p$ with a $10~M_J$ planet orbiting at $100$~au.
The horseshoe region, primary and secondary inner spiral arms, and an azimuthally elongated vortex can be clearly seen.
Dust depletion takes place in the inner disk ($\sim$40--50~au for $a=30~\mu$m, $\sim$30--50~au for $a=300~\mu$m) and it is more significant for larger dust particles because they suffer stronger drag.
The inner dust cavity grows over time, and at the end of the calculation ($t=100~T_p$) particles with $a \geq1$~mm are completely cleared from the inner disk.
The outer vortex is prominent in our simulations, but not seen in the ALMA sub-millimeter thermal emission map presumably because the actual SAO~206462 disk has a faster decline in surface density with radius than in our model at large radii \citep[$\gtrsim 100$~au;][]{andrews11}: our model disk has more mass in the outer disk than the observational estimates for the actual disk.
It is also possible that a slow dust growth rate at large radii limits dust mass with $a\gtrsim100~\mu$m in the outer disk of SAO~206462, which is responsible for the sub-millimeter thermal emission.

The width of the horseshoe region ($x_{\rm hs}$) is about 25~au, which is in good agreement with a known relation between the local disk aspect ratio $h = H/R$ and the planet-to-star mass ratio $q=M_p/M_*$: $x_{\rm hs} \sim R_p \sqrt{q/h}$ \citep{masset06,baruteau08,paardekooper08}.

\begin{figure}
\centering
\epsscale{1.15}
\plotone{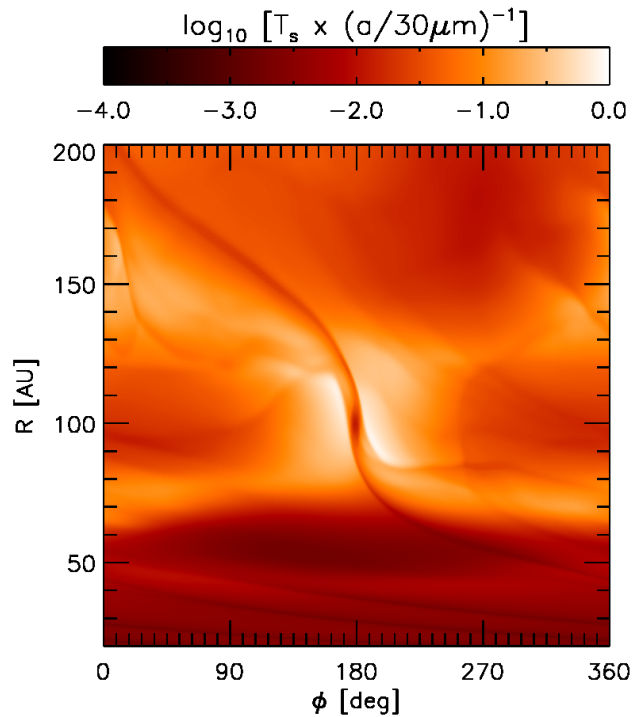}
\caption{Distribution of the Stokes number $T_s$ corresponding to the gas distribution shown in Figure \ref{fig:simulation}. Note that the numbers given in the colorbar are for the particles with $a=30~\mu$m.}
\label{fig:stokes}
\end{figure}

\begin{figure*}
\centering
\epsscale{1.15}
\plotone{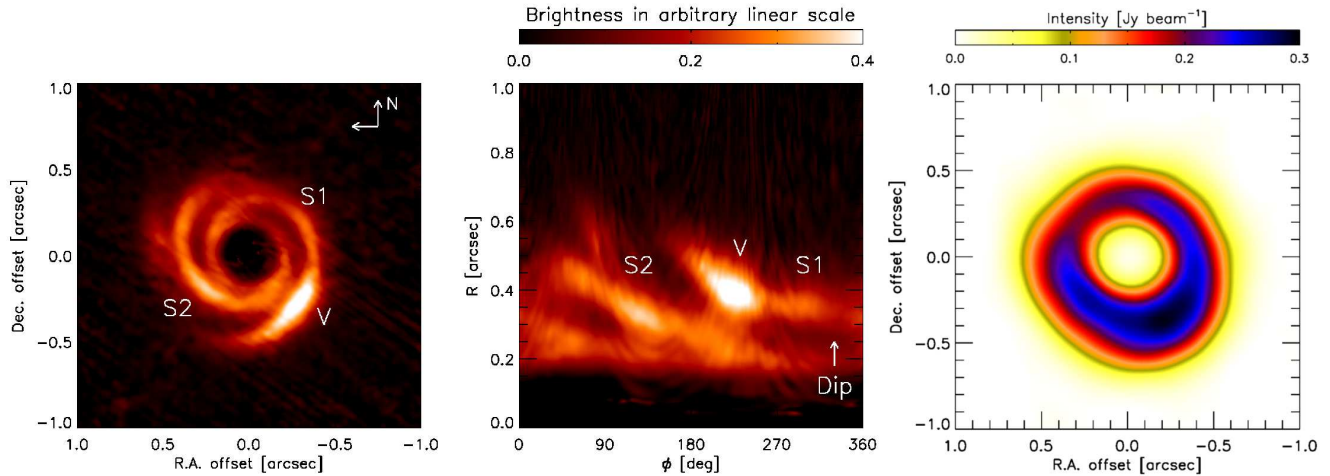}
\caption{(Left) Polarized scattered light observed in the K$_s$ band by \citet{garufi13}. The brightness is scaled with $R^2$ to compensate for stellar light dilution, and is in the arbitrary linear scale. We used publicly available data from the Vizier online catalog (http://vizier.cfa.harvard.edu/viz-bin/VizieR?-source=J/A+A/560/A105).
(Middle) Same as the left panel but in $\phi - R$ coordinates where $\phi$ is defined as the angle measured counter-clockwise from the north.
(Right) Dust continuum emission obtained with ALMA at 690~GHz by \citet{perez14}. We used publicly available data from the ALMA archive.
In the left and middle panels, we label some important structures: S2 refers to the arm on the eastern side, S1 refers to the arm that extends from the north to west, and V refers to the brightest peak on the southwestern side.}
\label{fig:obs}
\end{figure*}

The vortex efficiently traps dust particles, particularly the ones with a stopping time of the order of  the local orbital time as expected in typical anticyclonic vortices \citep[e.g.][]{barge95,inaba06,meheut12,fu14b,zhustone14,bae15}.
As a result the dust-to-gas mass ratio inside the vortex becomes as large as $\sim1:25$.
In Figure \ref{fig:stokes} we present the distribution of the Stokes number $T_s$ that corresponds to the model presented in Figure \ref{fig:simulation}.
The Stokes number in the vortex is about $10^{-3}$--$10^{-2}$ for the $30~\mu$m particles and is close to unity for millimeter-sized particles.
As the Stokes number increases against the particle size ($T_s \propto a$), larger particles exhibit a more compact structure.
For the $300~\mu$m-sized dust, the vortex extends $\sim180^\circ$  in azimuth.
The total (gas+dust) mass in the inner vortex varies over time, but it is about $3~M_J$ at $50~T_p$, which is roughly consistent to the mass constrained by sub-millimeter observation \citep[$2~M_J$;][]{perez14}.

Regarding the inner arms, we find that the secondary arm becomes significantly fainter outside of $\sim 60$~au, whereas the primary arm extends further out to the planet.
The faint secondary arm at $\gtrsim 60$~au may be due to the fact that less material exists because of the gap opened by the planet.
However, this may also be because the second spiral arm is excited at $m=2$ Lindblad resonance, which resides at $R \sim 0.63~R_p$. 
We note that the pitch angle of the primary arm increases as a function of radius, especially near the planet, while that of the secondary arm is nearly constant over radius.
It is also worth pointing out that the vortex and the inner spiral arms rotate at different frequencies: the vortex orbits at the local Keplerian speed whereas the spiral arms corotate with the planet.

\section{DISCUSSION}
\label{sec:discussion}

In Figure \ref{fig:obs}, we display the scattered light image at the K$_s$ band \citep{garufi13}, and the dust continuum emission at 690~GHz \citep{perez14}.
In the scattered light image, the disk shows two well-established spiral arms; one on the western side (hereafter S1) and the other on the eastern side (hereafter S2).
It also shows a bright azimuthal feature (hereafter V) at $(\Delta {\rm R.A.}, \Delta {\rm decl.}) = (-0.25'', -0.3'')$.
Interestingly, the bright feature in S1 in the scattered light is spatially coincident to the southwestern dust emission peak.

When the scattered light is plotted in $\phi - R$ coordinates, however, the pitch angle of the western arm changes abruptly in between S1 and V.
This raised the following question:
do S1 and V form a single spiral arm and share a common physical origin?
Or do they have different physical origins, but are just spatially coincident?

\begin{figure*}
\centering
\epsscale{1.15}
\plotone{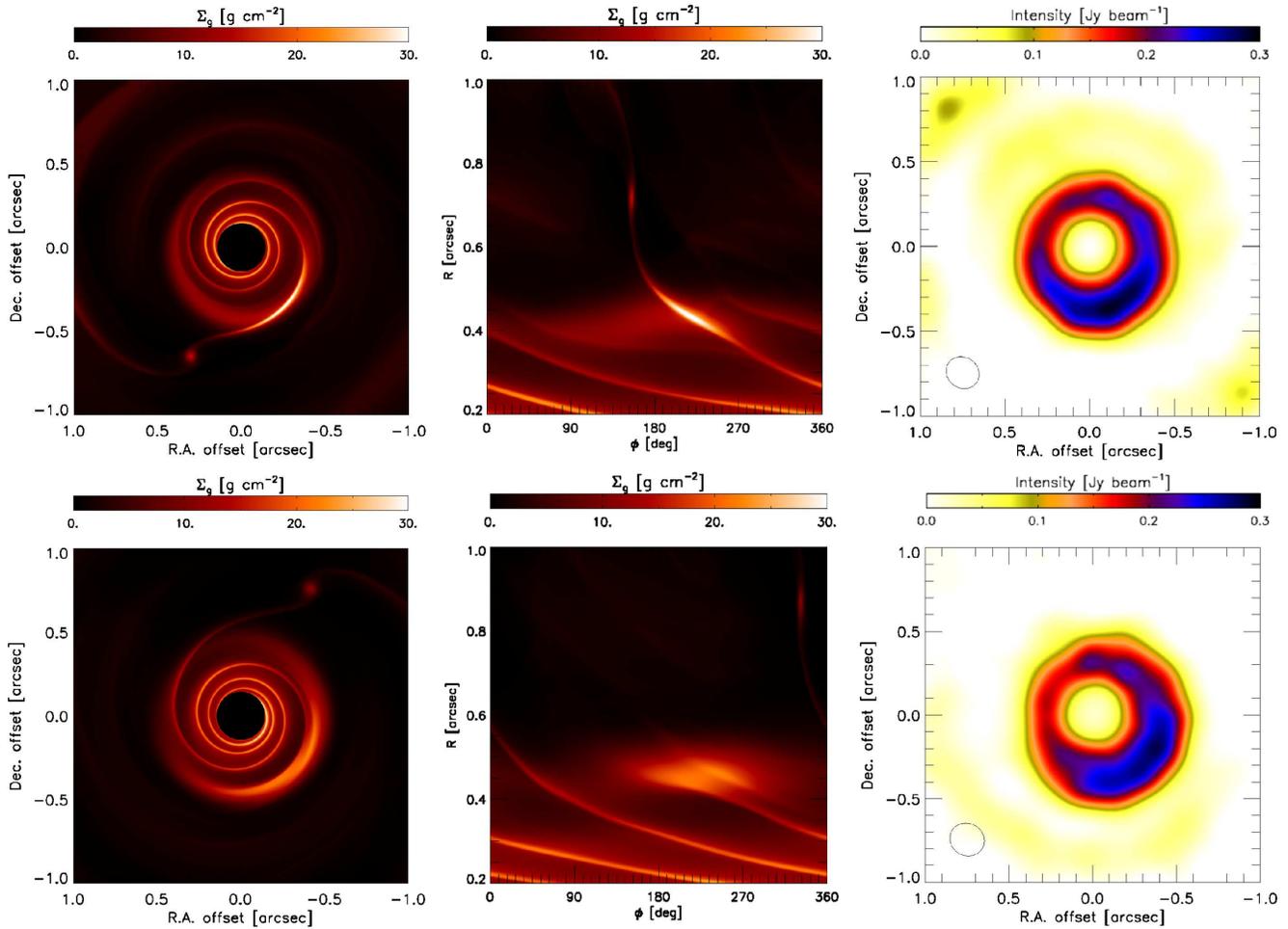}
\caption{{Upper panels:} when S1 is assumed to be the primary arm. $M_p = 10~M_J$ and $R_p = 100$~au were used. (Left) Gas density distribution in the inner 140~au disk (=1$''$) from this work. The planet is located at $(\Delta{\rm R.A.}, \Delta{\rm decl.}) = (0.''3, -0.''65)$. (Middle) Same as the left panel, but in $\phi-R$ coordinates where $\phi$ is measured counter-clockwise from the north. The primary and secondary inner spiral arms as well as the vortex are clearly seen. (Right) ALMA simulated image with the cycle 0 extended configuration. The synthesized beam is displayed in the lower-left corner.
{Lower panels:} same as the upper panels, but when S2 is the primary arm. $M_p = 15~M_J$ and $R_p = 120$~au were used. The planet is located at $(\Delta{\rm R.A.}, \Delta{\rm decl.}) = (-0.''4, 0.''75)$ in the left panel.}
\label{fig:model}
\end{figure*}

In order to infer the origin of the structures and to estimate the planet's mass and location, we compare the simulated surface density distribution and synthetic ALMA observations to the observed data.
We note that comparison between the simulated surface density distribution and the scattered light image should be treated with caution for several reasons.

\begin{figure*}
\centering
\epsscale{1.1}
\plotone{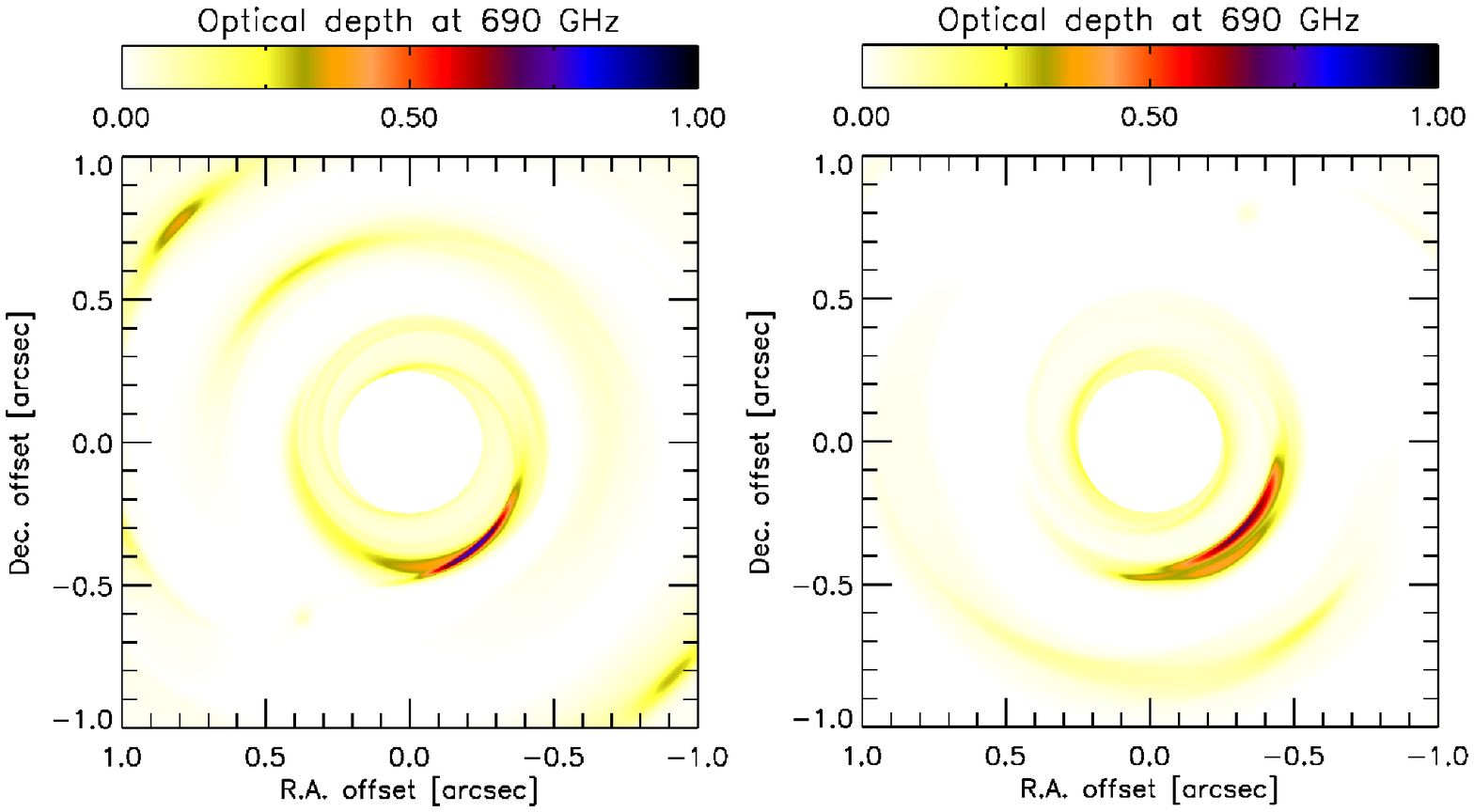}
\caption{Distribution of optical depth at 690~GHz for the two models presented in Figure \ref{fig:model}. We note that the entire disk is optically thin at this frequency, and only the core of the vortex is marginally optically thick. }
\label{fig:tau}
\end{figure*}

The near-infrared scattered light only traces the fluctuations of the disk atmosphere since the disk is highly optically thick at these wavelengths. 
Furthermore, the spiral shocks complicate the three-dimensional structure \citep{zhu15}.
Therefore, it is not straightforward to directly relate the disk surface density to the scattered light images.   
However, the shape of the spiral arms are similar at different heights in the disk (Figure 2 of \citealt{zhu15}) and similar to the spiral shape in the modeled scattered light images (Figure 2 of \citealt{fung15}).
Thus, we compare the morphology of the arms and density distribution in a qualitative manner, rather than in a quantitative manner.

The synthetic ALMA observations are produced as follows.
We use the Mie theory to calculate the dust opacity at 690~GHz. 
For particle sizes of 30, 100, 300, 1000, 3000~$\mu$m, the dust opacities are 12.0, 12.9, 7.6, 2.3, 0.6~${\rm cm}^2 {\rm g}^{-1}$, respectively.
For particles smaller than $17~\mu$m, we use the opacity at $a=3\mu$m as a representative value (2.4~${\rm cm}^2 {\rm g}^{-1}$).
We note that the resulting synthesized images are not sensitive to the choice of the opacity of $a<17~\mu$m particles because the mass fraction for the particles are only $5.3~\%$ (see Section \ref{sec:dust}).
Then, we calculate the optical depth of the disk as
\begin{equation} 
\tau = S \sum_{i} W_i \Sigma_{d,i} \kappa_i,
\end{equation}
where $W_i$ is the mass fraction ($0 \leq W_i \leq 1$, $\sum W_i = 1$) of each dust size bin, and $\Sigma_{d,i}$ and $\kappa_i$ are the dust surface density and the opacity for each dust size bin, respectively.
The scaling factor $S$ is introduced to calibrate the overall dust-to-gas ratio in a way that the simulated thermal emission matches to the observed intensity.
In the models introduced below, we use $S \sim 2$, which implies that the dust-to-gas mass ratio is presumably smaller than the one assumed in our calculations (1:100).
The total flux distribution is calculated with the temperature profile given in Equation (\ref{eqn:disk_temperature}), assuming the blackbody radiation with the effect of optical depth taken into account: $F = B_\nu(T) \exp(1-\tau)$, where $F$ is the flux, $B_\nu$ denotes the Planck function, and $\tau$ is the optical depth.
Finally, the flux is synthesized with the ALMA beam using CASA\footnote{http://casa.nrao.edu}.
Thermal noise from the atmosphere and from the ALMA receivers is added by setting the {\it thermalnoise} option in the {\it simobserve} task to {\it tsys-atm}.
We use the ALMA cycle 0 extended configuration to compare our models to the data presented in \citet[][see also Figure 2 of this paper]{perez14}.
The distance of 140~pc \citep[][and references therein]{muller11}, PA and inclination of $63^\circ$ and $16^\circ$ \citep{vandermarel15b} were used.

For the entire duration of our simulations ($100~T_p\lesssim10^5$~year) $\mu$m-sized dust particles are not completely depleted from the inner disk, while sub-millimeter and millimeter continuum observations suggest a significant reduction of dust density in the inner cavity \citep{lyo11,vandermarel15a,vandermarel15b}.
This is probably because the duration of the simulations is much shorter than the actual age of the system, although we cannot rule out the existence of a second companion inside the cavity.
In order to reduce the excess dust emission from the inner disk, we decrease the dust density at the inner 35~au by a factor of 100 when we produce synthetic ALMA observations.

\citet{zhu15} and \citet{fung15} recently showed that the azimuthal separation of scattered light from the primary and secondary arms is a function of planet-to-star mass ratio $q$.
In addition, the azimuthal separation of the arms is generally smaller than $180^\circ$ unless $q$ becomes very large ($q > 0.01$; \citealt{fung15}); so the secondary is generally ahead of the primary at the same radius.
The observed azimuthal separation between S1 and S2 ($|\phi_{S1}-\phi_{S2}|$) in SAO~206462 at $R=0.3-0.''35$ largely varies from $160^\circ$ to $260^\circ$,\footnote{Since it is not clear whether or not V is part of S1 at this point, we measure the separation between S1 and S2.}, partly because of the dip due to the depolarization effect around the minor-axis \citep[see][]{garufi13}.
Given the difficulties discriminating the primary and secondary, we test two scenarios: (1) S1 being the primary and (2) S1 being the secondary.
In both scenarios, we find that a vortex has to be located at the southwestern side to explain the observed sub-millimeter dust emission peak.

In Figure \ref{fig:model}, we present the gas surface density distribution in the inner $1"$ and the simulated ALMA observation.
The upper panels show the first scenario in which S1 is assumed to be the primary arm, with $M_p = 10~M_J$ and $R_p = 100$~au.
Interestingly, the primary arm passes through the vortex at the inner gap edge. 
While relating surface density to scattered light has to be done with caution as pointed out earlier, it is plausible that a spiral passing through a denser region would also produce higher density at the scattering surface.
Also, because the primary arm generally extends further out and has a larger pitch angle than the secondary, the lopsided vortex-like structure can be naturally explained.
The large separation between the secondary and the primary arms in this scenario is still a question.

When S2 is assumed to be the primary arm, V and S1 in the scattered light probably have a different origin because it is unlikely that (1) the secondary arm extends further out than the primary and (2) the pitch angle of the secondary arm abruptly changes at large radii, though interaction with a vortex may produce an unexpected outcome. 
In order to reproduce the lopsided dust emission, a larger planetary mass of $M_p = 15~M_J$ is assumed so that the planet generates a stronger pressure bump at the gap edge.
Also, because a more massive planet opens a gap further away $R_p = 120$~au is used: recall that the vortex forms at the inner gap edge whose position is a function of planetary mass.
Since there is no boost via interaction between the spiral and vortex as in the other scenario, significant vertical motion inside the vortex will be required to explain the bright scattered light at the vortex position.
In fact, it is possible that a vortex has a three-dimensional vertical motion which can lift small particles to the disk atmosphere, potentially enhancing the scattered light intensity  \citep{meheut12}.

It is possible that the spiral arms generate strong shocks and thus produce a high temperature near the surface \citep{zhu15}. 
Regarding the sub-millimeter flux, however, high temperature near the surface would not provide significant flux given that the disk is optically thin at 690 GHz as seen in Figure \ref{fig:tau} and most of the mass is concentrated near the midplane.

\begin{figure*}[t]
\centering
\epsscale{1.1}
\plotone{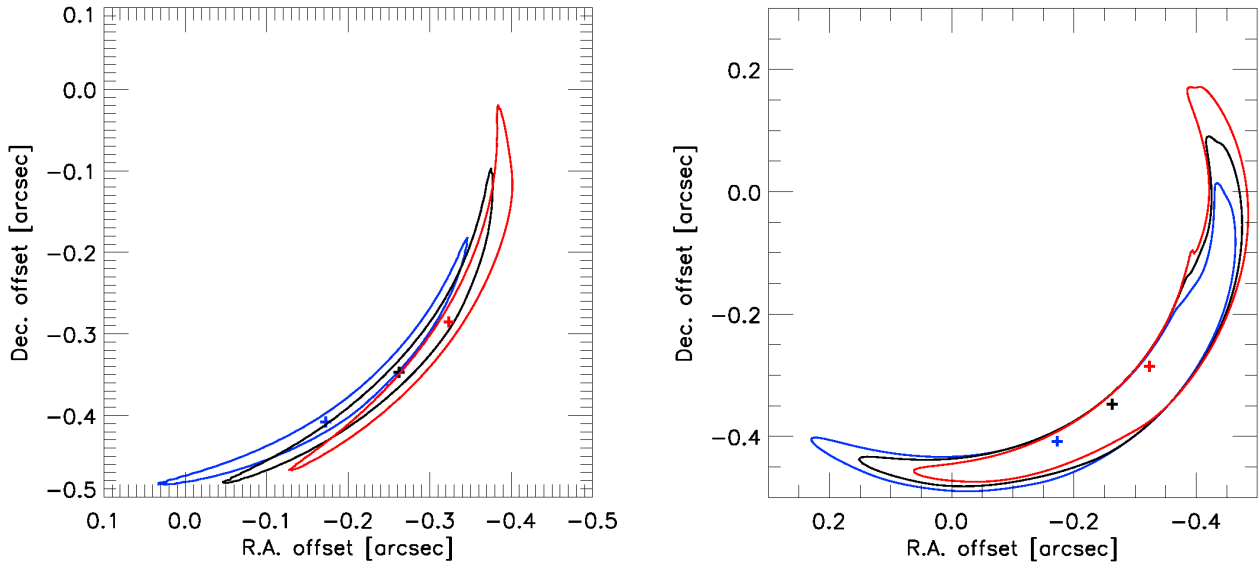}
\caption{Iso-density contours ($\Sigma_g = 20~{\rm g~cm^{-2}}$) showing time evolution of the bright southwestern feature in the two scenarios: (left) S1 is the primary arm and (right) S1 is the secondary arm. The black contours show positions at the same time as in Figure \ref{fig:model}. The blue and red contours show positions at $-10$ and $+10$ years from the black contours, respectively. If the bright feature originates from the vortex alone (scenario 2) it will move circularly over time, whereas the displacement will be circular + lateral if it is from a spiral arm passing through vortex (scenario 1). The cross symbols indicate the positions where the density peaks in the structures.}
\label{fig:prediction}
\end{figure*}

The vortex and the inner spiral arms not only orbit at different frequency as pointed out earlier, but they move in a different manner.
Figure \ref{fig:prediction} shows time evolution of the bright southwestern peak.
If the peak is originated from a spiral arm interacting with vortex it will appear to be more opened over time, showing both circular and lateral displacement.
On the other hand, if the peak is due to the vortex alone, it will move only circularly.
Assuming the vortex center is at $0.''4 = 56$~au, the peak will rotate about $1.^\circ1$ per year.
Since the two scenarios show noticeably different evolution, monitoring of the brightest peak over the next few years will help reveal the origin of the structures.

\citet{pinilla15} pointed out that the southwestern dust emission peak is shifted over time by comparing two ALMA data sets obtained in 2012 and 2014.
The shift is, however, much more significant than what would be expected from our models.
Instead, given that the two ALMA data sets are taken at different frequencies (690 and 340~GHz), it is possible that the observations show offsets among dust particles with different sizes because particles with Stokes numbers closer to unity are known to show more violent movement inside a vortex \citep[e.g.][]{bae15}.

We note that the observations so far were not able to reach the detection limit of $\sim 10~M_J$ at 100~au in SAO~206462 \citep[e.g.][]{vicente11}, because the thermal emissions from such planets are too small compared to the central star.
Future observations with GPI and SPHERE will be able to provide stronger constraints, if not detect a planet candidate.

Our calculations are limited to two dimensions and, therefore, we are unable to conclude which scenario would be preferred over the other at this point.
Future three-dimensional simulations and radiative transfer calculations will be able to provide a more robust prediction.
In addition, knowing more accurate estimates of the vortex position and disk temperature profile will help constrain the planetary mass and position for future searches.

\section{CONCLUSIONS}

We have constructed two-dimensional models with dust drift, which show vortex and spiral structure similar to that observed in SAO~206462.  
We further suggest that an interaction between a spiral arm and the vortex accounts for the abrupt change in brightness seen in the scattered light observations, although vortex alone may also explain the bright scattered light peak.  
Monitoring of the brightest peak over the next few years and higher resolution ALMA observations can help test our models and perhaps suggest which of our two preferred scenarios is more likely.

\acknowledgments

The authors thank the anonymous referee for a helpful report that improved the initial manuscript. 
The authors also thank Jeffrey Fung and Antonio Garufi for valuable discussion and Antonio Garufi et al. for making their data publicly available.
Z.Z. greatly appreciates Laurent Pueyo sharing his unpublished results and providing very helpful
suggestions.
This research was supported in part by the University of Michigan, and computational resources there provided by Advanced Research Computing.
Z.Z. is supported by NASA through Hubble Fellowship grants HST-HF-51333.01-A awarded by the Space Telescope Science Institute, which is operated by the Association of Universities for Research in Astronomy, Inc., for NASA, under contract NAS 5-26555. 
This paper makes use of the following ALMA data: ADS/JAO.ALMA $\#$2011.0.00724.S. ALMA is a partnership of ESO (representing its member states), NSF (USA) and NINS (Japan), together with NRC (Canada), NSC and ASIAA (Taiwan), and KASI (Republic of Korea), in cooperation with the Republic of Chile. The Joint ALMA Observatory is operated by ESO, AUI/NRAO and NAOJ. The National Radio Astronomy Observatory is a facility of the National Science Foundation operated under cooperative agreement by Associated Universities, Inc.

\appendix
\section{RESOLUTION TEST}

We double the resolution in order to check numerical convergency, with a $10~M_J$ planet orbiting at 100~au.
We test with a particle size of $a=300~\mu$m as a representative case because (1) dust evolution does not affect to gas evolution in our calculations and thus the gas structure remains identical no matter which dust particle size is used, and (2) the dust stopping time often limits the time step in two-fluid calculations, especially when a small dust particle size is assumed.
For instance, when $a=30~\mu$m is assumed, the dust stopping time is locally more than three orders of magnitude shorter than the orbital time (see Figure \ref{fig:stokes}), and therefore causes a very small time step.

\begin{figure*}
\centering
\epsscale{1.15}
\plotone{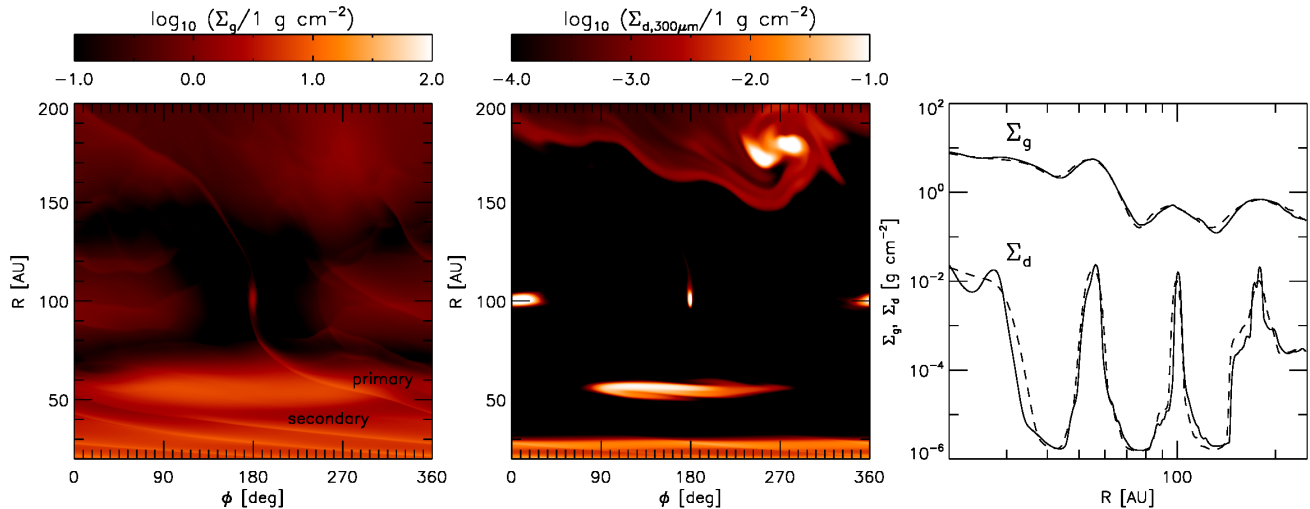}
\caption{Distributions of (left) gas, (middle) $300~\mu m$ dust in $\phi-R$ coordinates, and (right) azimuthally averaged radial profiles of the gas and dust distributions at $t=50.6~T_p$. The same model parameters are used as in Figure \ref{fig:simulation}, but with twice higher resolution ($n_R \times n_\phi$) = ($576\times1376$). In the right panel, the solid curves show high-resolution ($576\times1376$) results and the dashed curves show low-resolution ($288\times688$) results.}
\label{fig:resolution}
\end{figure*}

In Figure \ref{fig:resolution}, we display two-dimensional distributions of gas and $300~\mu$m dust, and their azimuthally averaged radial profiles.
While the high-resolution run shows more small-scale structures, the overall morphology is in good agreement with the low-resolution run presented in Figure \ref{fig:simulation}. 
The azimuthally averaged radial density distributions are also consistent with each other.

\end{document}